\documentclass[aps,pra,twocolumn,superscriptaddress]{revtex4-1}

\usepackage{graphicx}
\usepackage{amsmath,amssymb,amsfonts}
\usepackage{braket}
\usepackage{subfigure}
\usepackage{xcolor}
\usepackage{epstopdf}

\newcommand{\PT}{{\mathcal PT}}
\newcommand{\gth}{\gamma_{\mathrm{th}}}

\newcommand{\beqa}{\begin{eqnarray}}
\newcommand{\eeqa}{\end{eqnarray}}

\begin{document}
\title{Raising the $\mathcal{PT}$ transition threshold by strong coupling to neutral chains}

\author{Kaustubh S. Agarwal}
\affiliation{Department of Physics, Indiana University Purdue University Indianapolis (IUPUI), Indianapolis, Indiana 46202, USA}
\author{Rajeev K. Pathak}
\affiliation{Department of Physics, Savitribai Phule Pune University (SPPU), Ganeshkhind, Pune 411007, India}
\author{Yogesh N. Joglekar}
\affiliation{Department of Physics, Indiana University Purdue University Indianapolis (IUPUI), Indianapolis, Indiana 46202, USA}
\date{\today}

\begin{abstract}
The $\mathcal{PT}$ symmetry breaking threshold in discrete realizations of systems with balanced gain and loss is determined by the effective coupling between the gain and loss sites. In one dimensional chains, this threshold is maximum when the two sites are closest to each other or the farthest. We investigate the fate of this threshold in the presence of parallel, strongly coupled, Hermitian (neutral) chains, and find that it is increased by a factor proportional to the number of neutral chains. We present numerical results and analytical arguments for this enhancement. We then consider the effects of adding neutral sites to $\mathcal{PT}$ symmetric dimer and trimer configurations and show that the threshold is more than doubled, or tripled by their presence. Our results provide a surprising way to engineer the $\mathcal{PT}$ threshold in experimentally accessible samples.
\end{abstract}
\maketitle


\section{Introduction}
\label{sec:intro}
Over the past five years, parity-time ($\mathcal{PT}$) symmetric systems have been extensively investigated with experimental realizations in classical optics, electrical circuits, flying atoms~\cite{feng2017,el2018}, quantum photonics~\cite{xiao2017}, and ultracold atoms~\cite{luo2016}. A parity-time symmetric system is described by a non-Hermitian Hamiltonian that is invariant under the combined operations of parity ($\mathcal{P}$) and time reversal ($\mathcal{T}$); physically, it represents an open system with balanced, spatially separated, gain and loss~\cite{joglekar2013}

First discovered by Bender and Boettcher for continuum models on an infinite line~\cite{bender1998}, a $\mathcal{PT}$-symmetric Hamiltonian typically consists of a kinetic term $H_0=H_0^\dagger$ and a gain-loss potential term $\Gamma=-\Gamma^\dagger$, such that the total Hamiltonian $H_{PT}=H_0+\Gamma$ commutes with the antilinear $\mathcal{PT}$ operator. When the non-Hermiticity is small, the spectrum of $H_{PT}$ is purely real, and its eigenfunctions are simultaneous eigenfunctions of the $\mathcal{PT}$ operator with eigenvalue one. As the non-Hermiticity increases, two (or more) real eigenvalues become degenerate as do the corresponding eigenvectors. This exceptional point~\cite{kato2013} is called the $\mathcal{PT}$-symmetry breaking threshold. At this point, because the algebraic multiplicity of the relevant eigenvalue is larger than its geometric multiplicity~\cite{kunze}, the geometric eigenfunctions of $H_{PT}$ at the threshold do not span the entire space, and one can use the algebraic eigenfunctions to supplement them to form a basis~\cite{guntherrotter}. When the non-Hermiticity exceeds the $\mathcal{PT}$ breaking threshold, the spectrum is rendered into complex conjugate pairs. Thus, increasing the gain loss strength drives the open system from a quasi-equilibrium state ($\mathcal{PT}$-symmetric phase) to a state far removed from equilibrium ($\mathcal{PT}$-symmetry broken phase). 

Although a wide variety of lattice~\cite{jin2009,jin2010,jin2011,Joglekar2011a,Scott2011a,Vemuri2011,Liang2014,znoj2007a,znoj2007b,znoj2011} and continuum~\cite{bend2007,most2010}, linear and nonlinear~\cite{kono2016} $\mathcal{PT}$ symmetric models have been theoretically investigated, the experiments are limited to dimer (two sites)~\cite{rute2010,feng2014,hoda2014} or trimer (three sites)~\cite{hoda2017} realizations with balanced gain and loss. In these models, the $\mathcal{PT}$ breaking threshold is proportional to the tunneling amplitude and is determined by the overlap of evanescent fields between the two sites. Therefore, sweeping the threshold, say, over a decade is difficult. In principle, increasing the distance between the gain and loss sites reduces the $\mathcal{PT}$ breaking threshold exponentially; in practice, disorder effects become important when the tunneling amplitude becomes very small. The threshold can be reduced at will by using a time-periodic gain and loss with the right modulation frequency~\cite{jogl2014,lee2015}. But engineering a balanced, Floquet gain and loss is a challenging task. In the same vein, the $\mathcal{PT}$ transition threshold is, in principle, increased by making the gain and loss sites closer to each other. In practice, the different physical mechanisms necessary to implement loss in one and gain in the other put serious constraints on the minimum separation between the two sites. 

In this article, we present a new method to increase the $\mathcal{PT}$ symmetry breaking threshold. By  strongly coupling the gain-loss chain to a large number of neutral chains, {\it we show that the resultant threshold is increased by a factor equal to half the total number of coupled chains.} The plan of the paper is as follows. In Sec.~\ref{sec:model}, we introduce the tight-binding model used to describe the two-dimensional system under consideration, and recall the results for $\mathcal{PT}$ breaking threshold in a single chain~\cite{jogl2010}. Sec.~\ref{sec:num} starts with numerically obtained $\mathcal{PT}$ phase diagrams as a function of the number of coupled chains, the location of the $\mathcal{PT}$ symmetric chain, and the location of gain potential within a chain. Then we present a heuristic, analytical method for estimating the $\mathcal{PT}$ symmetry breaking threshold, and derive the threshold scaling law. In Sec.~\ref{sec:small} we consider the effect of surrounding $\mathcal{PT}$ symmetric dimers and trimers by neutral sites, and show that the threshold can be tuned from zero to triple its value in experimentally realistic systems. We conclude with a brief discussion in Sec.~\ref{sec:disc}. 
  

\section{Tight-binding model}
\label{sec:model}

\begin{figure}[b]
\centering
\includegraphics[width=\columnwidth]{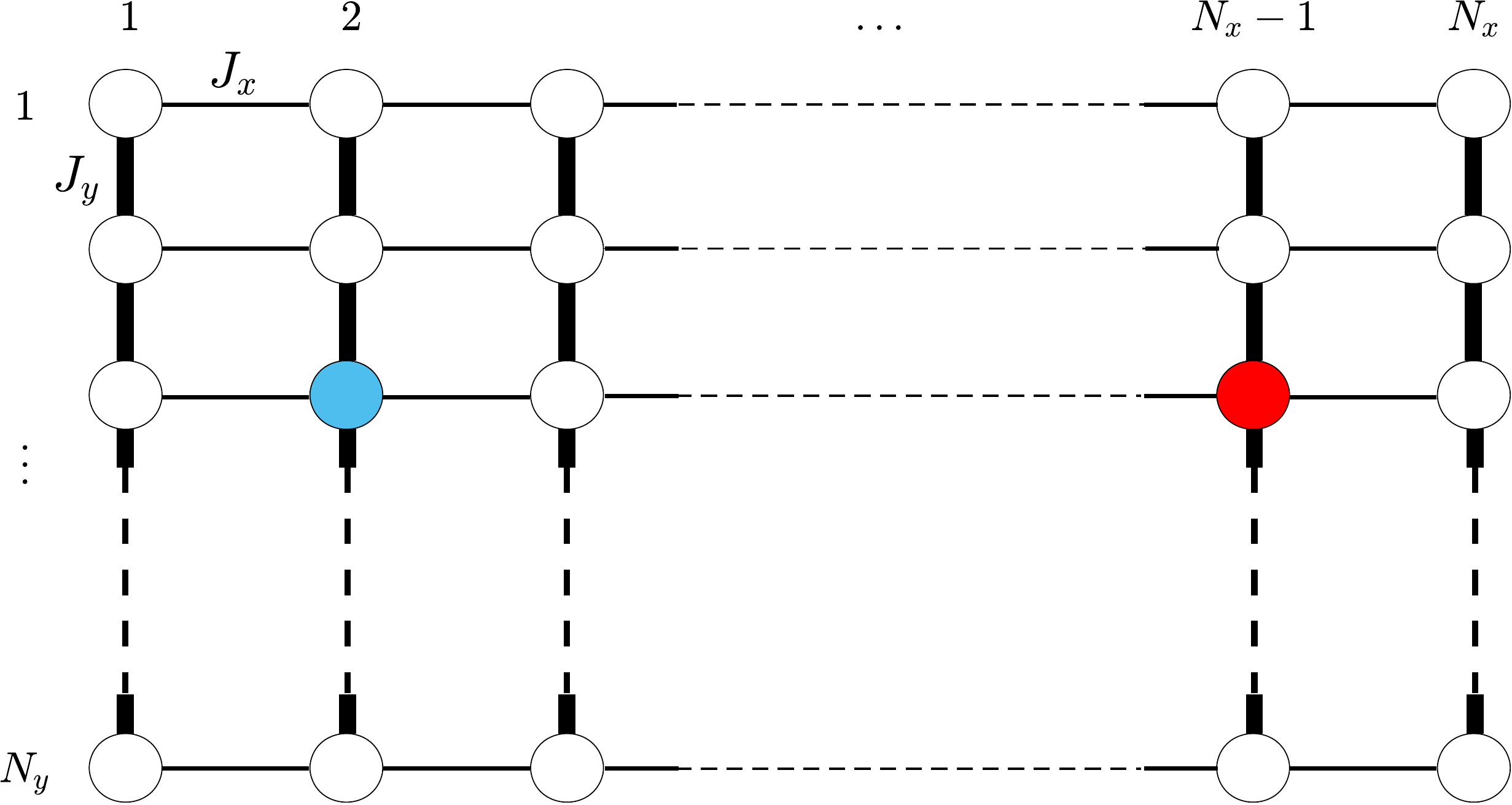}
\caption{Schematic of a $\mathcal{PT}$ symmetric chain with $N_x$ sites, strongly coupled to $N_y-1$ Hermitian chains of the same length. The gain site (blue) has a potential $+i\gamma$ and its parity-symmetric site (red) has the loss potential $-i\gamma$. The coupling within a chain is $J_x$ (thin black lines) and the strong, inter-chain coupling is $J_y\gg J_x$ (thick black lines). The sites in this $N_x\times N_y$ lattice are labeled by coordinates $(m,n)$ with $1\leq m\leq N_x$ and $1\leq n\leq N_y$.}
\label{fig:2dlattice}
\end{figure}

Consider a two-dimensional, finite, tight binding lattice with $N_x$ sites along the $x$-direction, $N_y$ sites along the $y$-direction, and open boundary conditions (Fig.~\ref{fig:2dlattice}). $J_x$ and $J_y$ denote the nearest-neighbor couplings along the two directions respectively. One of the horizontal chains has a gain potential $+i\gamma$, shown in blue, at location $(m_0,n_0)$ and a loss potential $-i\gamma$, shown in red, at its reflection-symmetric location $(\bar{m}_0,n_0)$ with $\bar{m_0}=N_x+1-m_0$. The Hamiltonian for this system is $H_{PT}=H_0+\Gamma$, where the Hermitian tight-binding part and the anti-Hermitian gain-loss part are given by 
\begin{eqnarray}
H_0& = & -J_x \sum_{m,n} \left(\ket{m,n}\bra{m+1,n} +\mathrm{h.c.}\right),\nonumber\\ 
  &  & -J_y \sum_{m,n}\left(\ket{m,n}\bra{m,n+1} + \mathrm{h.c.}\right),\label{eq:h0}\\ 
\Gamma & =  & +i\gamma\left(\ket{m_0,n_0}\bra{m_0,n_0} - \ket{\bar{m}_0,n_0}\bra{\bar{m}_0,n_0}\right),\label{eq:Gamma}
\end{eqnarray}
where $\ket{m,n}$ denotes a state localized at lattice site $(m,n)$ and h.c. denotes the Hermitian conjugate. The Hamiltonian $H_{PT}$ commutes with the $\mathcal{PT}$ operator where the action of the parity operator is given by $\mathcal{P}:(m,n)\rightarrow(\bar{m},n)$ and $\mathcal{T}=*$ is complex conjugation. The $\mathcal{PT}$-symmetry breaking threshold $\gth(m_0,n_0)$ of the Hamiltonian depends upon the eigenvalues and eigenvectors of $H_0(J_x,J_y)$. These are given by
\begin{eqnarray}
\label{eq:wavef}
\Psi _{p,q}(m,n) & \equiv & \braket{m,n|k_p,k_q} = A\sin(k_p m)\sin(k_q n),\\
\label{eq:energyf}
E_{p,q} & =& -2J_x\cos k_p -2J_y \cos k_q,
\end{eqnarray}
where $k_{p} = p\pi/(N_x + 1)$ and $k_{q} = q\pi/(N_y + 1)$ are the dimensionless quasimomenta consistent with open boundary conditions, $1\leq p\leq N_x$, $1\leq q\leq N_y$, and the normalization constant is given by $A=2/\sqrt{(N_x + 1)(N_y + 1)}$. 

When the horizontal chains are weakly coupled, $J_y/J_x\rightarrow 0$, the spectrum $E_{p,q}$ has $N_x$ energy levels each with a degeneracy of $N_y$. In this case, the $\mathcal{PT}$ threshold $\gth(m_0)$ shows a U-shaped behavior~\cite{jogl2010}. For even lattices, the $\mathcal{PT}$ symmetric phase is robust with $\gth=J_x$ when the gain and loss sites are the farthest, i.e. $m_0=1$, or the closest, i.e. $m_0=N_x/2$. For intermediate gain locations, the $\mathcal{PT}$ threshold algebraically goes to zero as the lattice size $N_x$ is increased~\cite{jogl2012,Joglekar2014a}. For odd lattices, the same behavior is true, except $\gth\rightarrow J_x/2$ when the distance between the gain and loss potentials is the  smallest, i.e. $m_0=(N_x-1)/2$~\cite{jogl2010}. 

When $J_y=J_x$, the threshold is suppressed to zero when the gain-loss Hamiltonian $\Gamma$ connects states that are degenerate due to the four-fold symmetry of the resultant square lattice~\cite{Ge2014}. In this article, we focus on the  strongly coupled chains, i.e. $J_y/J_x\gg 1$. In this limit, the spectrum in Eq.(\ref{eq:energyf}) has $N_y$ energy bands, with each band comprising $N_x$ eigenvalues spread over a width $\sim 4J_x$. Therefore, in the following, we will use the label $p$ to denote the level index within a band and $q$ to denote the band index. This separation of the spectrum into bands and levels within a band is valid when the bands do not overlap, and we will consider chains where this criterion is satisfied.


\section{$\mathcal{PT}$ symmetry breaking threshold results} 
\label{sec:num}

In this section we will present numerical results for the $\mathcal{PT}$ phase diagrams of coupled chains, followed by analytical derivation of the salient results.


\subsection{Numerical results}
\label{subsec:nr}

\begin{figure*}
\includegraphics[width=1.03\columnwidth]{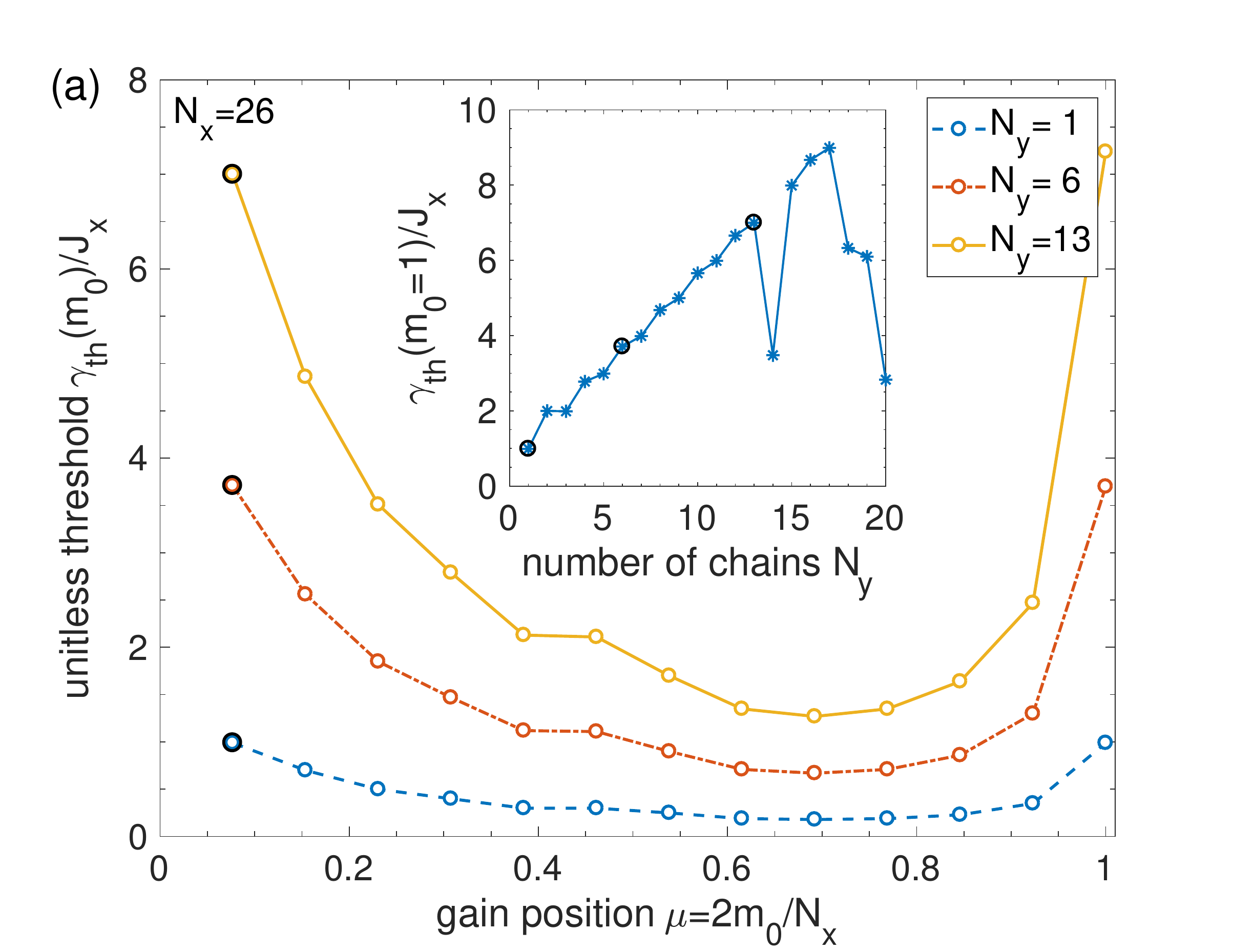}
\includegraphics[width=1.03\columnwidth]{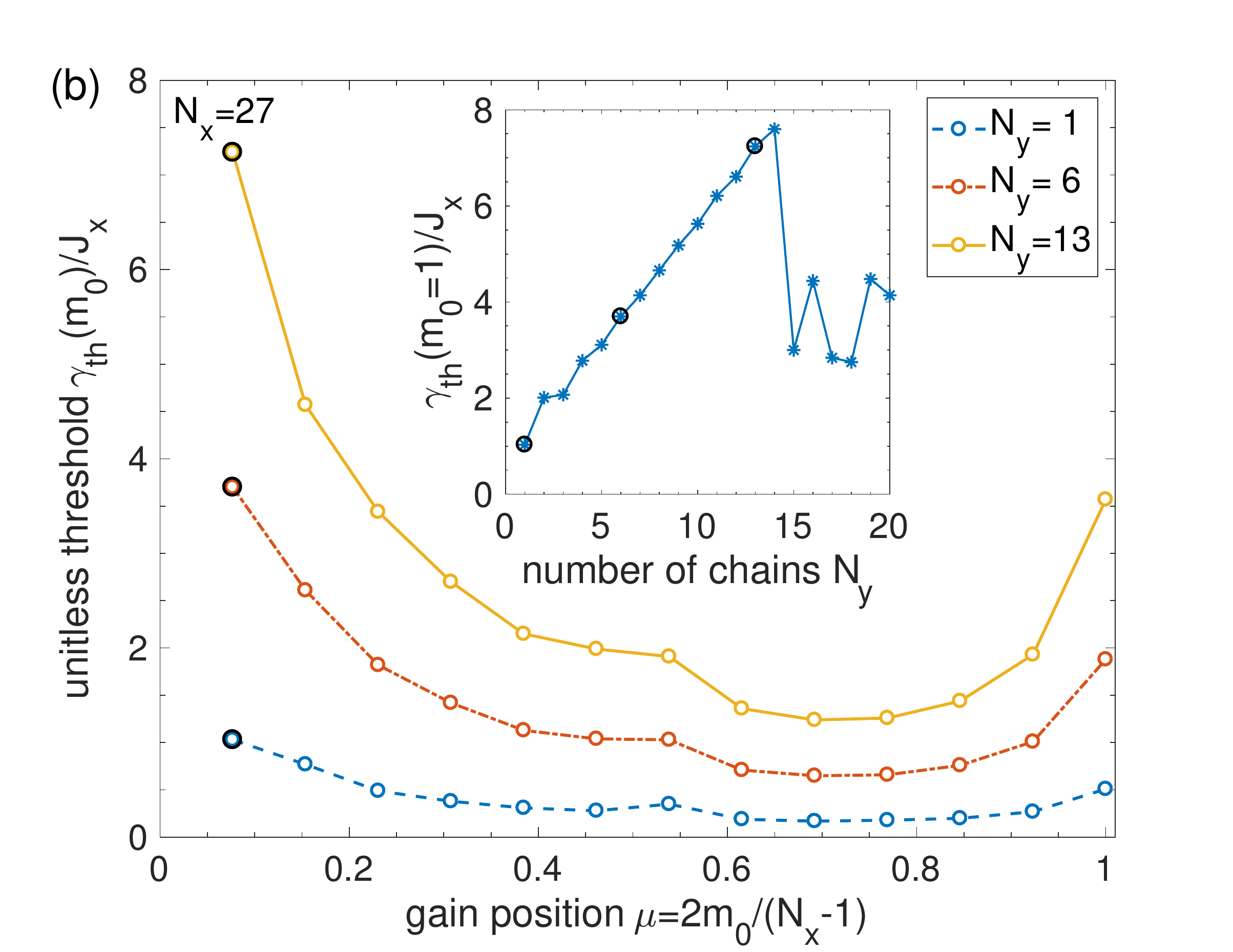}
\caption{Dependence of the $\mathcal{PT}$-breaking threshold $\gth(m_0)/J_x$ on the relative gain position $\mu=2m_0/N_x$ (or $\mu=2m_0/(N_x-1)$) and the number of horizontal chains $N_y$ for (a) even lattice with $N_x=26$ and (b) odd lattice with $N_x=27$. The couplings are $J_y/J_x=20$. For a single chain, $\gth(m_0)$ shows the characteristic U-shape. As $N_y$ is increased, the $\mathcal{PT}$-threshold increases as well. The insets in (a) and (b) show that the maximum value of the threshold, found for $m_0=1$, increases linearly with the number of chains. }
\label{fig:phaseopen}
\end{figure*}
We start with numerically obtained results for $\PT$-symmetry breaking threshold $\gth(m_0,n_0)$ as a function of the number of strongly coupled chains. Figure~\ref{fig:phaseopen} shows the behavior of $\gth$ as a function of the relative gain position $\mu=2m_0/N_x$ when the gain site is on the top chain, i.e. $n_0=1$. These results are for $J_y/J_x=20$. Panel (a) shows the results for an even chain with $N_x=26$ sites. When the number of chains is $N_y=1$, the threshold shows the characteristic U-pattern as a function of location of the gain site~\cite{jogl2010}. When the number of chains increases to $N_y=\{6,13\}$, the maximum value of the threshold increases monotonically with it. We remind the reader that this maximum occurs when the gain-loss potentials are farthest apart, i.e. $m_0=1$, or nearest neighbors, i.e. $m_0=N_x/2$. The inset shows that the maximum threshold $\gth(m_0=1)/J_x$ scales linearly with the number of horizontal chains $N_y$ up to a point, $N_y\lesssim 15$. These results are valid for all strongly coupled chains with an even number of lattice sites.

Figure~\ref{fig:phaseopen}(b) shows the results for an odd, $N_x=27$ lattice with $J_y/J_x=20$ and relative gain position $\mu=2m_0/(N_x-1)$. For a single chain, the threshold $\gth(m_0)$ shows the characteristic U-shape where the threshold for the nearest gain-loss location, $m_0=(N_x-1)/2$, is half of that for the farthest gain and loss, $m_0=1$~\cite{jogl2010}.  As the number of chains $N_y$ is increased, the threshold $\gth(m_0)$ increases in a proportionate manner. The inset shows the linear dependence of the largest threshold $\gth(m_0=1)/J_x$ on the number of chains. When $N_y\gtrsim 14$ this linear relationship breaks down, as it does in Fig.~\ref{fig:phaseopen}(a). These results are valid for all strongly coupled chains with an odd number of lattice sites. 

\begin{figure}
\centering
\includegraphics[width=\columnwidth]{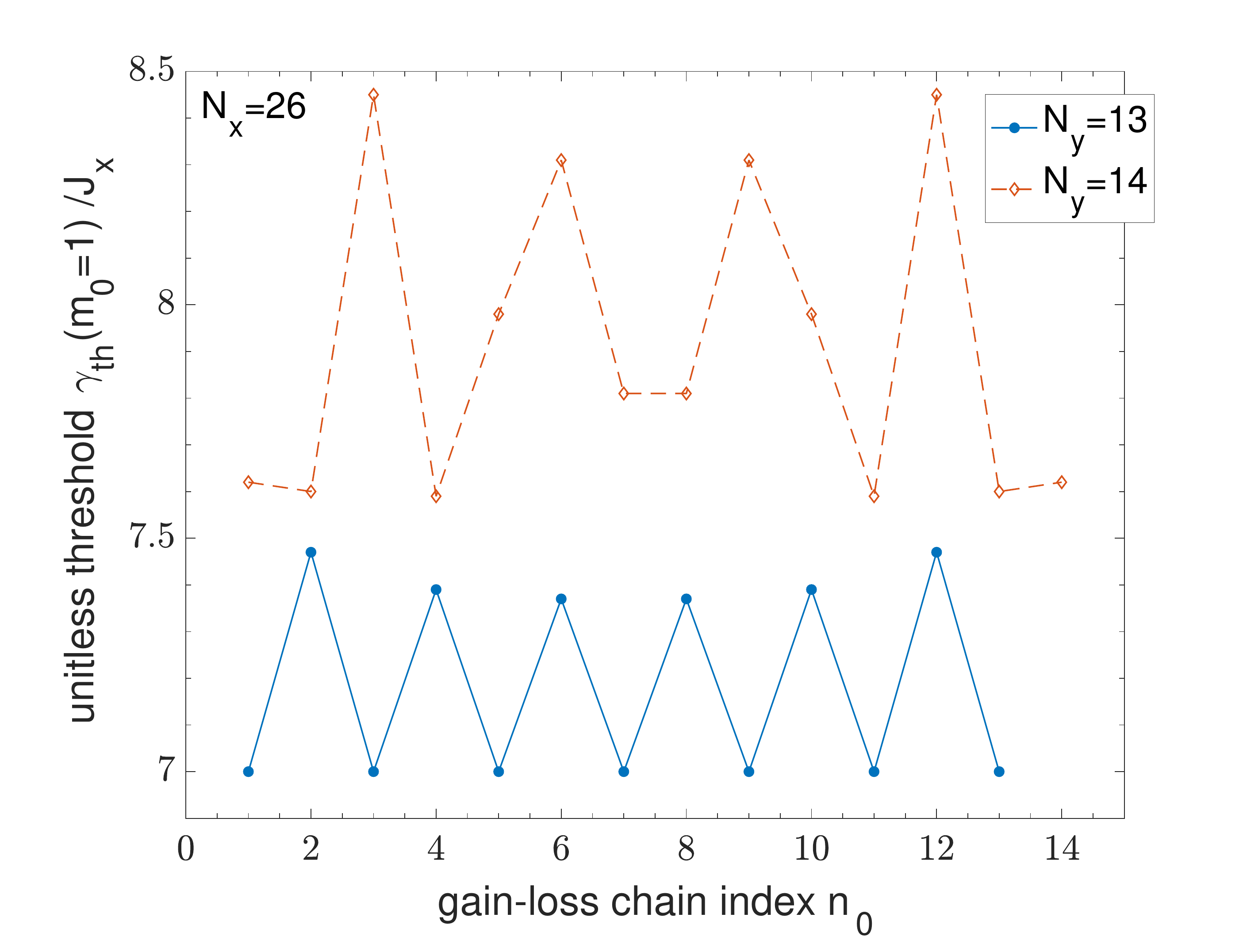}
\caption{Minor variations of the $\mathcal{PT}$ breaking threshold $\gth(m_0=1)/J_x$ as a function of the chain index $n_0$ for a system with $N_y=13$ (solid blue) and $N_y=14$ (dashed red) chains, each with $N_x=26$ sites. The coupling strength is $J_y/J_x=30$.}
\label{fig:Nx26Nyvsn0}
\end{figure}

The results presented in Fig.~\ref{fig:phaseopen} are for a configuration when all the neutral chains are on one side of the $\mathcal{PT}$ symmetric chain. How do they change when the $\mathcal{PT}$ symmetric chain is embedded within the parallel neutral chains? Figure~\ref{fig:Nx26Nyvsn0} shows that when its location is changed from the top ($n_0=1$) to halfway down ($n_0=(N_y+1)/2$ or $n_0=N_y/2$), the threshold remains roughly constant. 

\begin{figure*}
\includegraphics[width=1.03\columnwidth]{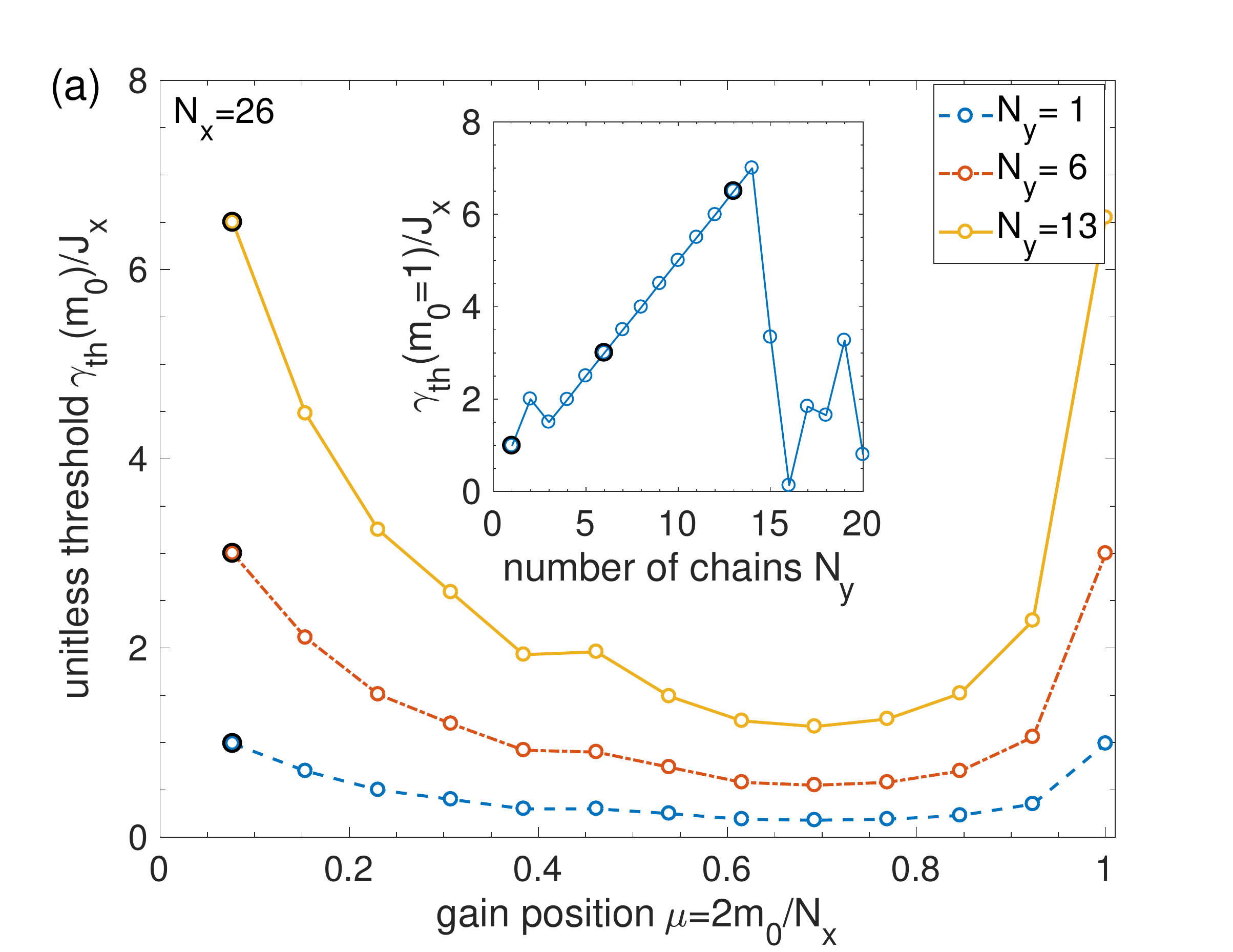}
\includegraphics[width=1.03\columnwidth]{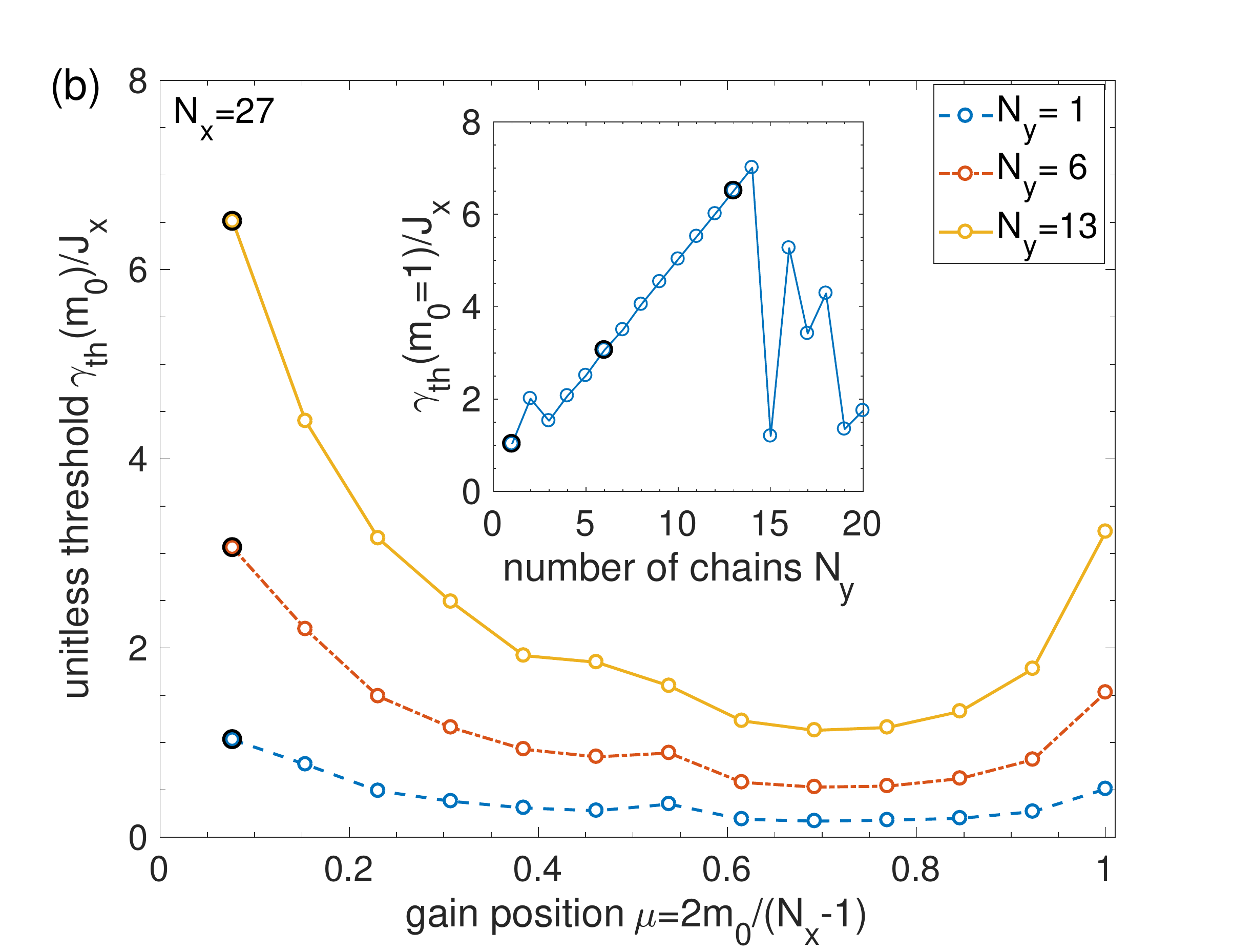}
\caption{Dependence of the $\mathcal{PT}$-breaking threshold $\gth(m_0)/J_x$ on the number of horizontal chains $N_y$ with periodic boundary conditions along the $y$-direction for (a) an even, $N_x=26$ lattice and (b) an odd, $N_x=27$ lattice. As $N_y$ is increased, the $\mathcal{PT}$-threshold increases while maintaining a characteristic U-shaped profile. The insets in (a) and (b) show that the threshold  $\gth(m_0)$ increases linearly with $N_y$.}
\label{fig:phasepbc}
\end{figure*}
This insensitivity of the $\mathcal{PT}$ symmetry breaking threshold $\gth(m_0,n_0)$ to the location of the gain-loss chain is also borne out by the phase diagrams for a system with periodic boundary conditions along the $y$-direction. When the system is periodic in $N_y$, the threshold $\gth$ will be independent of the index $n_0$ of the gain-loss chain.  Figure~\ref{fig:phasepbc}(a) shows the results for an $N_x=26$ site chain with $J_y/J_x=20$ as a function of gain location $m_0$ for increasing numbers of coupled chains. The inset shows that the threshold for $m_0=1$ scales linearly with $N_y$ for periodic boundary conditions along the $y$-direction. Figure~\ref{fig:phasepbc}(b) shows that the scaling law remains valid for odd-sized chains ($N_x=27$) as well. We remind the reader that when periodic boundary conditions are imposed along the $x$-direction, due to the resultant degeneracies in the spectrum of the Hermitian Hamiltonian $H_0$, the threshold is reduced to zero irrespective of the gain-loss distance, i.e. $\gth(m_0)=0$ for all $m_0$.~\cite{scot2012}.

Note that in Figs.~\ref{fig:phaseopen} and~\ref{fig:phasepbc}, the inset shows that the linear-in-$N_y$ scaling of the $\mathcal{PT}$ threshold breaks down as $N_y$ is increased. This breakdown is due to the finite values of $J_y/J_x$ used in the calculations. We will show in the next subsection that the scaling is only valid in the ``strong coupling'' regime that is defined by Eq.(\ref{eq:strongcoupling}). Thus, for a fixed value of $J_y/J_x$ as the number of coupled chains is increased, the system ceases to be in the strongly coupled regime. 

The salient finding from Figs.~\ref{fig:phaseopen},~\ref{fig:phasepbc}, and~\ref{fig:phasepbc} is that for a large number of chains $N_y\gg 1$, in the strong coupling limit, the $\mathcal{PT}$ breaking threshold for $N_y$ coupled chains is strongly renormalized. For open boundary conditions, Fig.~\ref{fig:phaseopen}, we get
\begin{equation}
\label{eq:gth2d}
\lim_{N_y\gg 1}\gamma_{\mathrm{th}}(m_0,N_y)=\left(\frac{N_y+1}{2}\right)\gamma_{\mathrm{th}}(m_0,N_y=1),
\end{equation}
whereas for periodic boundary conditions along the $y$-direction, the scaling factor is $N_y/2$ instead of $(N_y+1)/2$. Since the threshold $\gth(m_0)$ is algebraically fragile~\cite{jogl2010,jogl2012,Joglekar2014a} for all gain locations except $m_0=1$ or $m_0=N_x/2$, we have chosen $m_0=1$ for the results shown in Figs.~\ref{fig:phaseopen} and~\ref{fig:Nx26Nyvsn0}. A similar scaling behavior is also obtained when the gain-loss potentials are closest to each other. In the following paragraphs, we will present a heuristic, analytical derivation of this result. 


\subsection{Derivation of the threshold scaling law}
\label{subsec:ar}

The $\mathcal{PT}$ symmetry breaking threshold is determined by the $\gamma$-flow of two (or more) energy eigenvalues that develop level attraction and become degenerate as the gain-loss strength approaches the threshold value. In a large lattice with strong anisotropy, i.e. $N_x,N_y\gg 1$ and $J_y/J_x\gg 1$, the eigenvalues $E_{p,q}$, Eq.(\ref{eq:energyf}), of the Hermitian Hamiltonian $H_0$ are divided into $N_y$ subbands, each of which has $N_x$ energy levels. The ground-state energy is $E_{1,1}\sim -2J_y-2J_x$, the first subband is characterized by energy levels $E_{p,1}$ with $1\leq p\leq N_x$, and it ends at $E_{N_x,1}\sim -2J_y+2J_x$. The second subband starts at $E_{1,2}$ and goes up to $E_{N_x,2}$. These bands are well separated from each other provided the lowest level in the $(q+1)$th band, i.e $E_{1,q+1}$, is higher than the highest level of the $q$th band, i.e. $E_{N_x,q}$, for all $q \in \{1,\dots ,N_y\}$. The cosine-band structure has the highest density of states at the bottom of the band, and thus, the smallest gap between adjacent subbands occurs when $q=1$. As a result, the criterion for well-separated bands is given by $E_{N_x,1}<E_{1,2}$ and reduces to 
\begin{equation}
\label{eq:strongcoupling}
\frac{J_y}{J_x}>\left[\frac{\cos(\frac{\pi}{N_x + 1}) - \cos(\frac{N_x\pi}{N_x + 1})}{\cos(\frac{\pi}{N_y + 1}) - \cos(\frac{2\pi}{N_y + 1})}\right].
\end{equation}
The right-hand side in Eq.(\ref{eq:strongcoupling}) reduces to $4N_y^2/3\pi ^2$ in the limit $N_x,N_y\gg 1$. Thus, the results we obtain in the following paragraphs are valid in the limit $N_x\gg 1$, $N_y\gg 1$, and $J_y/J_x\gg N_y^2$. On the other hand, when these criteria are not met, we expect that the  scaling law, Eq.(\ref{eq:gth2d}), will break down (as seen in Figs.~\ref{fig:phaseopen} and~\ref{fig:phasepbc}). 

To obtain the adjacent levels $(p,p+1)$ within a subband $q$ that will drive the $\mathcal{PT}$-symmetry breaking transition we use the following procedure. Consider the Hamiltonian $H_{PT}$ in the $2\times2$ subspace spanned by orthonormal states $|p,q\rangle$ and $|p+1,q\rangle$.  Apart from a constant energy-shift term that we can safely ignore, the effective Hamiltonian is given by 
\begin{equation}
\label{eq:Heff}
H_{\mathrm{eff}}(m_0,n_0)=(E_{p,q}-E_{p+1,q})\frac{\sigma_z}{2}+i\Delta_{p,q}(m_0,n_0)\sigma_x,
\end{equation}
where $\sigma_z$ and $\sigma_x$ are the Pauli matrices. The effective potential $\Delta$ is 
obtained by taking the matrix elements of the gain loss potential $\Gamma$, Eq.(\ref{eq:Gamma}), in the basis of the two eigenstates $|p,q\rangle$ and $|p+1,q\rangle$, 
\begin{eqnarray}
i\Delta_{p,q} & \equiv & \langle p,q|\Gamma|p+1,q\rangle,\nonumber\\
& = & 2i\gamma A^2 \sin(k_{p}m_0)\sin(k_{p+1}m_0)\sin^2(k_q n_0).
\label{eq:delta}
\end{eqnarray}
It follows from Eq.(\ref{eq:Heff}) that the $\mathcal{PT}$ threshold for the effective $2\times 2$ model is determined by 
\begin{equation}
\label{eq:delta2}
|\Delta_{p,q}(\gamma_\mathrm{var})|=\frac{1}{2}|(E_{p,q}-E_{p+1,q})|.
\end{equation} 
Depending on the level index $p$, this gives rise to $(N_x-1)N_y$ different variational numbers $\gamma_{\mathrm{var}}(p,q)$. Since the $\mathcal{PT}$ symmetry is broken when it breaks for any pair of adjacent levels, we use the minimization of this variational threshold as the criterion for determining the level-index pair $(p_0,p_0+1)$ and the band index $q_0$.

\begin{figure*}
\centering
\includegraphics[width=\textwidth]{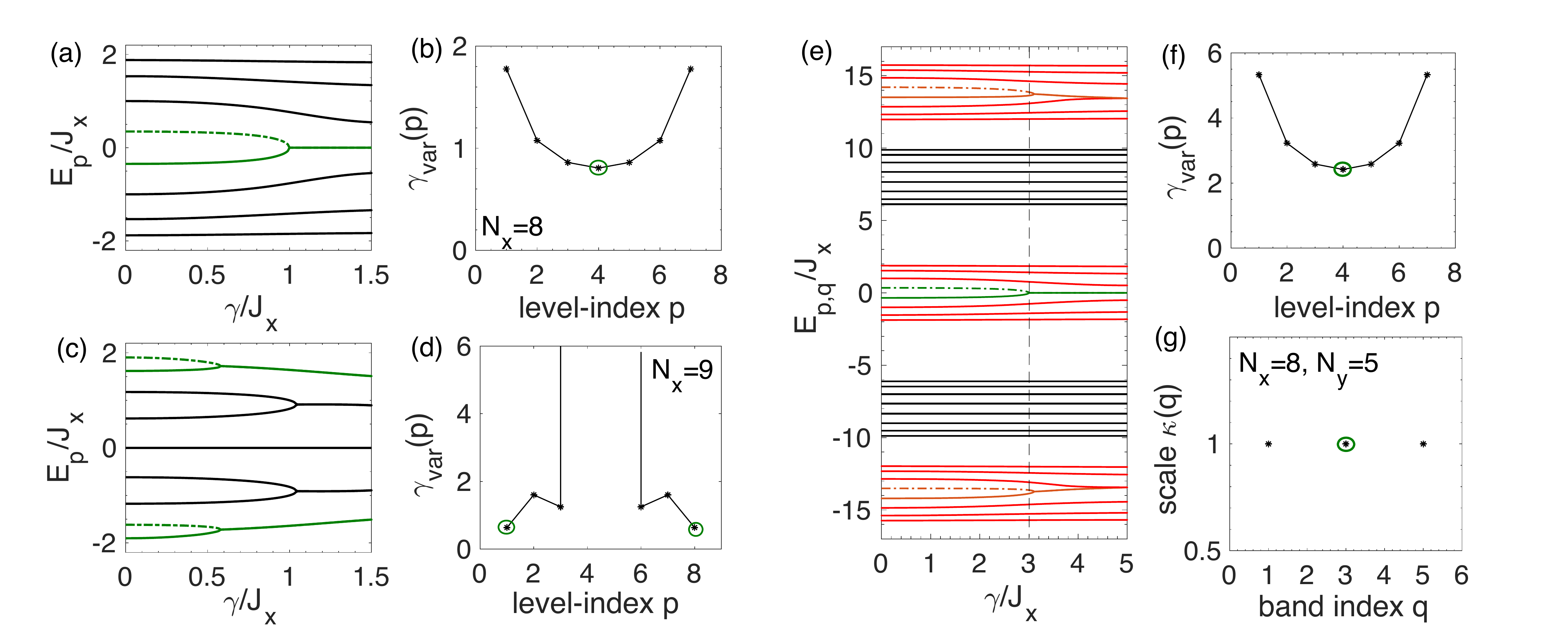}
\caption{Variational approach for determining the levels $(p,p+1)$ that break the $\mathcal{PT}$ symmetry in a single $N_x=8$ chain, (a)-(b), and $N_x=9$ chain, (c)-(d). (a) eigenvalue flow shows that the levels at the band center drive the $\mathcal{PT}$ transition when $m_0=1$. (b) variational $\gamma_{\mathrm{var}}(p)/J_x$ obtained from Eq.(\ref{eq:Heff}) predict that the $\mathcal{PT}$ breaking level is $p_0=4$. (c) eigenvalue flow for a chain with $m_0=(N_x-1)/2$ shows that the band-edge levels drive the transition. (d) variational approach for $\gamma_{\mathrm{var}}(p)/J_x$ gives the same result. The divergent $\gamma_{\mathrm{var}}$ for $p=\{4,5\}$ reflects the fact that the mid-gap state is  unaffected by $\gamma$. Panels (e)-(g) show results for an $8\times 5$ system with $J_y/J_x=8$ and gain location $(m_0,n_0)=(1,3)$. (e) eigenvalue flows shows that central levels, $p=4$, in the central band, $q=3$ are responsible for the transition. (f) shows that $\gamma_{\mathrm{var}}(p)$ has a minimum at $p=4$, and $q=3$ is the optimal band index for minimizing both $\gamma_{\mathrm{var}}$ and the scale factor $\kappa(q)$, panel (g).}
\label{fig:eflow}
\end{figure*}

Does this heuristic method work? Figure~\ref{fig:eflow} presents the results of such an analysis. For a single chain with $N_x=8$ levels and $m_0=1$, the flow of eigenvalues $E_p(\gamma)$ shows that the two, particle-hole symmetric levels at the band center drive the $\mathcal{PT}$ breaking transition, panel (a). Panel (b) shows that the variational threshold $\gamma_{\mathrm{var}}(p)$ obtained from $H_{\mathrm{eff}}$ reaches a minimum at level index $p=4$, matching with the results from panel (a). Panels (c)-(d) depict the corresponding results for an $N_x=9$ site chain with shortest distance between the gain and loss potentials. The variational threshold $\gamma_{\mathrm{var}}(p)$ is minimum at $p=1$ and $p=8$, which is consistent with the eigenvalue flow diagrams showing that particle-hole symmetric pairs of levels given by $(1,2)$ and $(8,9)$ driven the $\mathcal{PT}$ transition in this case. These representative results show that for one-dimensional chains, minimizing the variational threshold $\gamma_{\mathrm{var}}(p)$ for the $2\times 2$ Hamiltonian (\ref{eq:Heff}) accurately identifies the energy levels that drive the $\mathcal{PT}$-symmetry breaking transition. 

Panels (e)-(g) in Fig.~\ref{fig:eflow} show the results for a two-dimensional configuration of $N_y=5$ chains with $N_x=8$ sites, coupling ratio $J_y/J_x=8$, and gain location $(m_0,n_0)=(1,3)$. The eigenvalue flows in panel (e) show that the $\mathcal{PT}$ breaking occurs due to the central two levels of the central band at gain-loss strength $\gth/J_x=3=(N_y+1)/2$. Panel (f) shows that the variational threshold $\gamma_{\mathrm{var}}(p)$ has a minimum at $p=4$. The dependence of the scale factor 
\begin{equation}
\label{eq:scale}
\kappa(q)=\frac{\gamma_{\mathrm{var}}(p,q)}{\gamma_{\mathrm{var}}(p,q=1)}
\end{equation}
is shown in panel (g), where the divergent values of $\kappa$ at $q=\{2,4\}$ are not plotted. These results show that the ``minimization of the variational threshold'' strategy also works for strongly coupled chains in two dimensions.

It follows from the effective Hamiltonian $H_{\mathrm{eff}}$ that the scale factor $\kappa(q)$ for the optimal level index $p$ and band-index $q$ simplifies to 
\begin{equation}
\label{eq:scale2}
\kappa=\min_{q}\frac{(N_y+1)}{2}\mathrm{cosec}^2\left(\frac{q\pi n_0}{N_y+1}\right).
\end{equation}
This equation is obtained as follows. Eq.(\ref{eq:delta2}) determines the variational threshold $\gamma_{\mathrm{var}}(p,q)$ for a single chain and $N_y$ strongly coupled chains. The normalization factors $A^2$ for the two cases, however, are different. For a single chain, $A^2=2/(N_x+1)$, while that for $N_y$ strongly coupled chains is $A^2=4/(N_x+1)(N_y+1)$. This $(N_y+1)/2$-fold increase for a single chain, in essence, is instrumental to linear-in-$N_y$ scaling behavior. The cosecant term in Eq.(\ref{eq:scale2}) is bounded below by one, and for $N_y\gg 1$, it is always possible to choose a $q\in\{1,\ldots,N_y\}$ such that the argument of the cosecant-term is arbitrarily close to $\pi/2$. Therefore, the ratio of the two thresholds scales linearly with the number of horizontal chains as seen in Eq.(\ref{eq:gth2d}), i.e.
\begin{equation}
\lim_{\frac{qn_0\pi}{N_y+1}\rightarrow\frac{\pi}{2}}
\kappa=\frac{\gth(m_0,N_y)}{\gth(m_0,N_y=1)}\rightarrow\left(\frac{N_y+1}{2}\right).
\end{equation}
If the system is periodic along the $y$-direction, it is easy to check that the normalization factor for the eigenfunctions of the Hamiltonian $H_0$ changes to $A=2/\sqrt{(N_x+1)N_y}$. Therefore the scaling factor $\kappa$ changes to $N_y/2$ from $(N_y+1)/2$. 

Recall that the present derivation is based on the assumption of well separated bands, i.e. Eq.(\ref{eq:strongcoupling}). Therefore, we expect it to break down when $N_y\gtrsim\pi\sqrt{3J_y/4J_x}$. Indeed, insets in Figs.(\ref{fig:phaseopen}) and (\ref{fig:phasepbc}) show that the deviation from the linear behavior occurs around this value. As an aside, we note that for an odd-sized chain with the gain potential on the first site, $m_0=1$, three levels at the center of the band become degenerate at the $\mathcal{PT}$-symmetry breaking threshold and give rise to a third-order exceptional point~\cite{jogl2010,hoda2017}. However, since the zero-energy level remains unchanged across the $\mathcal{PT}$ symmetry breaking transition, we can restrict ourselves to the subspace of other two levels that change with the gain-loss strength. 


\begin{figure*}
\centering
\includegraphics[width=\textwidth]{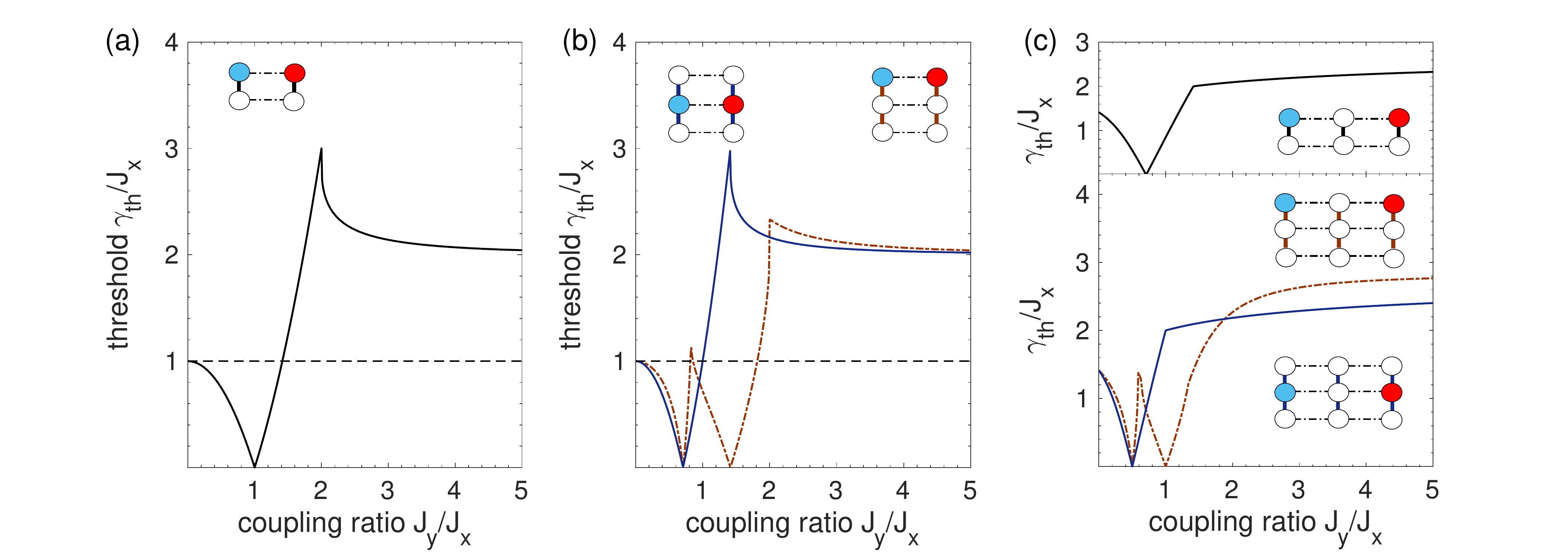}
\caption{$\mathcal{PT}$ threshold dependence on coupling to neutral sites shown by blank circles; the gain-site is shown in blue and the loss-site is shown in red. (a) $\gth/J_x$ for a $\mathcal{PT}$-neutral dimer system is tripled from its single-dimer value when $J_y/J_x=2$, and saturates to two in the strong coupling limit, $J_y/J_x\gg 1$. (b) for three dimers, the threshold is more than doubled near $J_y/J_x\sim 2$ and saturates to two for $J_y/J_x\gg 1$. (c) Corresponding results for a $\mathcal{PT}$-neutral trimer system (top panel), and a three-trimers system (bottom panel). In all cases, the threshold is more than doubled even at moderate values of $J_y/J_x\gtrsim 2$.}
\label{fig:small}
\end{figure*}
\section{$\mathcal{PT}$ dimer and trimer plaquettes}
\label{sec:small}

In the previous section, we discovered a strong growth of the $\mathcal{PT}$ symmetry breaking threshold in the limit of many, strongly coupled, long chains, i.e. $N_y\gg 1$, $J_y/J_x\gg N_y^2$, and $N_x\gg 1$. Motivated by ``the unreasonable effectiveness of mathematics in the natural sciences''~\cite{wigner}, we will now consider the applicability of those results to two chains, $N_y=2$, with two or three sites each, $N_x=\{2,3\}$, and small to moderate coupling ratio $J_y/J_x\sim O(1)$. Such configurations are nothing but dimer or trimer plaquettes; their symmetrical versions, $J_x=J_y$, have $\gth=0$ due to resultant discrete rotational symmetry and their nonlinear versions have been investigated in the past~\cite{li2012}. Here, we focus on the asymmetrical cases, i.e. $J_y\neq  J_x$, that are  experimentally realizable. 

First, let us consider a gain-loss dimer (strongly) connected to a neutral dimer (Fig.~\ref{fig:small}(a)). The $4\times 4$ Hamiltonian for such a system is given by 
\begin{equation}
\label{eq:h4}
H_4(\gamma)=(-J_x\sigma_x+i\gamma\sigma_z)\otimes{\bf 1}_2+{\bf 1}_2\otimes(- J_y\sigma_x).
\end{equation}
It is straightforward to obtain the particle-hole symmetric eigenvalues 
\begin{equation}
\label{eq:lambda4}
\lambda_k=\pm\left[J_x^2+J_y^2-\frac{\gamma^2}{2}\pm\frac{1}{2}\sqrt{\gamma^4+16 J_x^2J_y^2-4\gamma^2J_y^2}\right]^{1/2}.
\end{equation}
The $\mathcal{PT}$ transition threshold $\gth/J_x$ can be analytically obtained from Eq.(\ref{eq:h4}). Depending on the ratio $J_y/J_x$ the pair of eigenvalues, among the four given in Eq.(\ref{eq:lambda4}), that drive the $\mathcal{PT}$ breaking transition varies. This variation gives rise to the three distinct functional forms for the threshold function $\gth(J_y)$ seen in Fig.~\ref{fig:small}(a). In a similar spirit, we also consider a $\mathcal{PT}$ dimer connected to two neutral dimers, and a $\mathcal{PT}$ trimer connected to one or two neutral trimers. Figure~\ref{fig:small} shows the dependence of the threshold $\gth/J_x$ on the ratio of coupling strengths $J_y/J_x$ for different plaquette configurations. 

Figure~\ref{fig:small}(a) schematically shows a $\mathcal{PT}$-neutral dimer system. Starting from unity, the dimensionless threshold $\gth/J_x$ decreases to zero for the symmetrical configuration, i.e. $J_y=J_x$, but then rises rapidly to three when $J_y=2J_x$. As the asymmetry increases, $J_y/J_x\gg 1$, the threshold saturates to two.  Results for one $\mathcal{PT}$ dimer with two neutral dimers are shown in panel (b). When the $\mathcal{PT}$ dimer is in the middle, the threshold is first suppressed to zero, and then rises to three when $J_y/J_x=\sqrt{2}$ (solid blue line). In contrast, when the $\mathcal{PT}$ dimer is on top, the threshold vanishes at two different coupling strengths, and reaches a maximum near $J_y/Jx=2$ (dot-dash red line). In both configurations, the threshold saturates to $\gth/J_x=2$ in the strong coupling limit. 

Figure~\ref{fig:small}(c) shows the corresponding results for a $\mathcal{PT}$ trimer. When connected to another neutral trimer (top panel),  the threshold $\gth/J_x$ first decreases down to zero, then increases, and saturates to $\gth/J_x=2\sqrt{2}=2\gth(J_y=0)$. When we have two neutral trimers (bottom panel), the threshold shows a qualitatively similar behavior. The results in Fig.~\ref{fig:small} show that the $\mathcal{PT}$ transition threshold in experimentally realizable configurations is dramatically changed by coupling the $\mathcal{PT}$-dimer or $\mathcal{PT}$-trimer to neutral sites. 


\section{Discussion}
\label{sec:disc}
In this paper, we have studied the effects of surrounding a $\mathcal{PT}$ symmetric chain with neutral chains of the same length. The primary effect is that {\it the $\mathcal{PT}$ transition threshold is increased by a factor equal to half the total number of chains.} Although our analysis was carried out for many, long, strongly coupled chains, the results are also true for experimentally realizable $\mathcal{PT}$ symmetric dimers and trimers. The $\mathcal{PT}$ symmetry breaking thresholds in these systems are increased by a factor of two to three. 

$\mathcal{PT}$ symmetric models in two dimensions have not been extensively explored because, for most lattice or continuum models with rotational symmetries, the transition threshold is zero~\cite{Szameit2011,Ge2014,agar2015}. Our results show that highly asymmetrical, two-dimensional lattice models, with a ``few'' balanced gain and loss sites, give rise to a strong renormalization of the $\mathcal{PT}$ symmetry breaking threshold. Generalizing these results to other $\mathcal{PT}$ symmetric lattice model will provide deeper insights into these findings. 


\begin{acknowledgments}
This work was supported by NSF Grant no. DMR 1054020 (Y.J.) and R.K.P. gratefully acknowledges an honorary adjunct professorship by SPPU.
\end{acknowledgments}


%


\end{document}